\documentclass[english,aps,manuscript]{revtex4}
\usepackage[T1]{fontenc}
\usepackage[latin9]{inputenc}
\usepackage{amsthm}
\usepackage{amsmath}
\usepackage{graphicx}
\usepackage{amssymb}
\usepackage{esint}

\makeatletter
\@ifundefined{textcolor}{}
{%
 \definecolor{BLACK}{gray}{0}
 \definecolor{WHITE}{gray}{1}
 \definecolor{RED}{rgb}{1,0,0}
 \definecolor{GREEN}{rgb}{0,1,0}
 \definecolor{BLUE}{rgb}{0,0,1}
 \definecolor{CYAN}{cmyk}{1,0,0,0}
 \definecolor{MAGENTA}{cmyk}{0,1,0,0}
 \definecolor{YELLOW}{cmyk}{0,0,1,0}
 }

\makeatother

\usepackage{babel}

\begin{document}

\title{A Lie-Algebraic Approach To the Kondo Problem}

\author{S. G. Rajeev}

\email{rajeev@pas.rochester.edu}

\affiliation{Department of Physics and Astronomy}

\address{Department of Mathematics,University of Rochester, Rochester NY 14627
USA}
\begin{abstract}
The Kondo problem is studied  using the unitary Lie algebra of spin-singlet
fermion bilinears. In the limit when the number of values of the spin
$N$ goes to infinity the theory approaches a classical limit, which
still requires a renormalization. We determine the ground state of
this renormalized theory. Then we construct a quantum theory around
this classical limit, which amounts to recovering the case of finite
$N$.
\end{abstract}
\maketitle

\section{Introduction}

Renormalizability-especially asymptotic freedom- is a deep and fundamental
property of four-dimensional non-abelian gauge theories\cite{Huang}.
Unfortunately, at the moment we understand these theories well only
in perturbation theory. Although qualitative connections with AdS
theories of gravity have been made\cite{AdSCFT}, a precise quantitative
solution for the spectrum of a non-abelian gauge theory remains out
of our grasp. If we could get a non-perturbative formulation of the
theory, free of divergences, we would be closer to the goal of proving
that there is a mass gap and string tension\cite{QYMT}.

The Kondo problem \cite{Hewson} provides us with a simpler,but quite
deep, example of renormalization-and asymptotic freedom- with its
own intrinsic physical interest. In a sense, it is the {}``hydrogen
atom'' of renormalization theory. It should be possible to solve
this basic problem by simple methods, without unwieldy numerical calculations
or clever ansatzes that depend on too many details of the particular
model. Once we understand how to do this, we can see how similar ideas
might apply to gauge theories and their dual gravities.

Electrons in a metal move more or less like free particles, occasionally
scattered by the ions which oscillate around their equilibrium positions
by thermal fluctuations. As the temperature decreases, the ions move
less and the resistance should decrease. It does, except that at some
low temperature ($T\sim10K$) the resistance starts to increase again,
rising to a finite value as $T\to0.$ Kondo's explanation was that
metals can have magnetic impurities (e.g., Iron atoms embedded within
a Copper lattice) whose magnetic moments become ordered at low temperatures.
These little magnets can scatter electrons too, an additional contribution
to resistance at low temperature.The size of an atomic impurity is
very small compared to the wavelength of the electron: it is much
like a $\delta$ function interaction. This causes divergences: Kondo
calculated the magnetic contribution to resistance to be proportional
to $\log T^{-1}.$\textbf{ }Thus Kondo explained why resistance grows
at low temperatures.Theflaw is that it predicts infinite resistance
at zero temperature, contradicting experimental observations. This
is the Kondo problem.

Wilson's landmark solution\cite{Wilson} follows a numerical approach
(the Numerical Renormalization Group). The essential complication
is that there are an infinite number of electrons, all of whom contribute
to the divergent part of the interaction. The ingenious methods Wilson
devised have not been generalized to the case of multiple impurities.
Also, they have not yet been made mathematically rigorous.

Andrei and Wiegmann solved \cite{Andrei,Wiegmann} the Kondo problem
using Bethe Ansatz methods.While very elegant mathematically (and
potentially rigorous), the method is not flexible enough to be useful
beyond the basic example of the single impurity and with a linear
dispersion relation : in essence it is a fiendishly clever guess that
happens to work because of the special symmetries of the problem.

Nozieres\cite{Nozieres} introduced a simple physical picture that
is quite appealing: the magnetic impurity forms a bound state with
an electron (or hole) and then mostly decouple from the rest of the
electrons, leading to a Fermi liquid. It would be great to recover
this from the more fundamental Kondo hamiltonian: somewhat analogous
to recovering the chiral model from QCD. Affleck and Ludwig \cite{AffleckLudwig}come
closest to recovering such a picture based on conformal field theory:
representations of Virasoro and Kac-Moody Lie Algebras. The motivation
of our current work is in part to extend their studies beyond the
critical point, to situations where conformal symmetry may not be
exact.

We work with a larger Lie algebra than Kac-Moody or Virasoro, spanned
by all the spin-singlet bilinears of the electrons: even those that
are not local field operators. We will see that these observables
have small fluctuations in the limit where the spins take a large
number of values $N$: the bilinears have commutators of order$\frac{\hbar}{N}.$
Thus, even as $\hbar$ is kept fixed (for example, set equal to one)
the commutators can be approximated by classical Poisson brackets.
This {}``neo-classical'' limit\cite{Neoclassical} retains many
of the essential features of the theory, like the logarithmic divergence
of the coupling constant. We will be able to perform renormalizaton
explicitly and obtain the ground state in this limit. We then quantize
the excitations around this ground state, thus recovering the finite
$N$ theory. This strategy was applied earlier to two dimensional
QCD\cite{QCD2}, (which is free of UV divergences) as well as the
Chiral Gross-Neveu model\cite{HendersonRajeev}(also called the non-abelian
Thirring model). The Kondo problem is in fact simpler than either
of these cases. Nevertheless, we believe it is instructive to work
it out explicitly.

The large $N$ limit of the Kondo problem has been studied before\cite{Mora}.
But the usual diagrammatic approaches are not the best way to understand
how a new non-perturbative ground state forms. Our methods are more
similar to the variational principles of BCS theory: but they allow
for a systematic expansion around the neo-classical answer.

While there remain technical and mathematical details to be explained,
we recover a simple picture: an impurity-electron condensate forms,
and the excitations around this ground state are electron-like quasi-particles,
but with a modified spectrum of energies.

There has been a revival of interest in the Kondo effect because it
occurs in quantum dots. Since the parameters can be tuned\cite{QDot},
this experimental realization holds promise of testing the theory
as well as the potential for the invention of new devices based on
the Kondo effect.

\section{Fermion Bilinears}

Define fermionic operators satisfying the Canonical Anti-Commutation
Relations,

\[
[A^{\dagger k\sigma},A_{l\sigma'}]_{+}=\delta_{l}^{k}\delta_{\sigma'}^{\sigma}\]

\[
[A^{\dagger d\sigma,}A_{d\sigma'}]_{+}=\delta_{\sigma'}^{\sigma}\]

all other pairs of anti-commutators being zero. Throughout this paper
(even where we speak of classical dynamics and Poisson brackets),
we will use units such that $\hbar=1;$ in particular, it is not equal
to zero. Here, $k$ labels the momentum of a conduction band electron,$d$
an impurity state, and $\sigma$ the spin. Although $\sigma$ takes
just two values in the real system, it will be convenient to let it
take $N$ values. There is only impurity, $d=1$ . (We hope to generalize
later to the case where several impurities are present.) Capital letters
$K,L$ will denote indices that can take either conduction band or
impurity values: $K=k$ or $d.$ The momenta take a finite
range of values $k=-\Lambda,\cdots,-1,1,\cdots\Lambda$. The interesting
physical region is when the energies are small compared to $\Lambda,$or
equivalently, $\Lambda\to\infty.$ But this limit is very subtle,
requiring a renormalization of a coupling constant. Our strategy will
be to understand this first in the case of large $N,$ and only then
pass to the case of finite $N.$

Define the spin-zero bilinears \[
\Phi_{L}^{K}=\frac{1}{N}A^{\dagger K\sigma}A_{L\sigma}.\]

It is straightforward to check the commutation relations (the idea
that fermion bilinears span a unitary Lie algebra go back to Schwinger
in the early days of quantum field theory.)

\[
\left[\Phi_{L}^{K},\Phi_{N}^{M}\right]=\frac{1}{N}\left[\delta_{L}^{M}\Phi_{N}^{K}-\delta_{N}^{K}\Phi_{L}^{M}\right].\]

Thus, in the limit of large $N,$ these commutators become small:
the quantum fluctuations in the spin-zero bilinears are small. They
will tend to classical observables, the commutators being replaced
by Poisson Brackets. We can think of the finite $N$ case as the quantization
of these Poisson Brackets, with $\frac{1}{N}$ playing the role of
$\hbar$ in the usual quantum theory. This approach to the large $N$
limit is motivated by the theory of solitons in the theory of strong
interactions. In the large $N$ limit, the Heisenberg equations of
motion for the bilinears will tend to Hamiltonian equations. Such
unusual classical limits (where a classical theory emerges even as
$\hbar$ is finite, but some other parameter goes to zero) have been
called `neo-classical' in another context. We will solve for the static
solution of least energy. For the solution to remain well-defined
as $\Lambda\to\infty,$ we will have to renormalize a coupling constant
even in this neo-classical limit.

Then we have the P.B. in the large N limit

\[
-i\left\{ \Phi_{L}^{K},\Phi_{N}^{M}\right\} =\left[\delta_{L}^{M}\Phi_{N}^{K}-\delta_{N}^{K}\Phi_{L}^{M}\right].\]

\section{The Kondo Hamiltonian}

The Kondo hamiltonian is

\[
H=\sum_{k}\omega_{k}A^{\dagger k\sigma}A_{k\sigma}+JA^{\dagger d\sigma}A_{d\sigma}\sum_{k}A^{\dagger k\sigma}\sum_{k'}A_{k'\sigma'}.\]

The sum over all momentum of the conduction band electrons amounts
to evaluating the operators at the position of the impurity (the origin).

We have allowed the spin generators to form a $U(N)$ Lie algebra
rather than $SU(N):$ this is a minor change, as the added singlet
decouples from the rest. But it makes for easier book keeping.(In
't Hooft's study of the large $N$ limit of gauge theories, a similar
passage from $SU(N)$ to $U(N)$ gauge group is made.)

Of interest is anti-ferromagnetic case  $J>0$, where the  electron and the impurity will form a spin-singlet.   Define the operators

\[
\Phi_{d}^{\bullet}=\sum_{m}\Phi_{d}^{m}\]

\[
\Phi_{\bullet}^{d}=\sum_{m}\Phi_{m}^{d}.\]

which  involve the mixing of a conduction band state and
a state located at the impurity. Expressed in terms of the bilinears,
the Kondo hamiltonian is

\[
H=\sum_{k}\omega_{k}\Phi_{k}^{k}-J\Phi_{d}^{\bullet}\Phi_{\bullet}^{d}\equiv H_{0}-JH_{1}\]
Assume that the energies of the conduction band electrons are $\omega_{k}$
are non-degenerate:

\[
\omega_{k}\neq\omega_{l},\quad\mathrm{if}\quad k\neq l\]

and that \[
\omega_{-k}=-\omega_{k}.\]
(Charge conjugation symmetry.) Moreover, we assume that the dispersion
relation is asymptotically linear: \[
\lim_{|k|\to\infty}\frac{\omega_{k}}{k}=c\]
for some constant $c.$ Unlike in the Bethe Ansatz method, our approach
does not rely on an exactly linear dispersion relation. Note that
we do not allow $\omega_{k}$or $k$ to take the value $0;$ this
is to avoid an annoying zero mode and is not an essential restriction.

\section{The Effective Hamiltonian}

Define the effective hamiltonian to be the matrix

\[
h_{L}^{K}(\Phi)=\frac{\partial H}{\partial\Phi_{K}^{L}}(\Phi).\]

Then

\[
h_{l}^{k}=\omega_{k}\delta_{l}^{k}\]

\begin{equation}
h_{d}^{k}=-g(\Phi),\quad h_{k}^{d}=-g^{*}(\Phi)\label{eq:h}\end{equation}

\[
h_{d}^{d}=0\]

where,

\[
g(\Phi)=J\Phi_{d}^{\bullet}.\]

Using the fact that the Poisson brackets are those of the Unitary
Lie algebra, it is easy to check that the equations of motion in the
large $N$ limit are, in matrix language,

\[
-i\frac{d\Phi}{dt}=[h(\Phi),\Phi].\]

Note, by the way, that $\Phi_{d}^{d}$ is a conserved quantity.

In particular, a static solution (such as the ground state ) will
satisfy

\[
[h(\Psi),\Psi]=0.\]

\section{Diagonalization Of The Effective Hamiltonian}

For a given value of $g,$ the eigenvalue equation for $h$ is,

\[
(\omega_{k}-\nu)U^{k}-gU^{d}=0\]

\[
-g^{*}U^{\bullet}=\nu U^{d}\]

So that

\[
(\omega_{k}-\nu)U^{k}+\frac{|g|^{2}}{\nu}U^{\bullet}=0\]

and

\[
U^{k}=|g|^{2}\frac{1}{\nu(\nu-\omega_{k})}U^{\bullet}.\]

Summing over $k,$the factor $u^{\bullet}$ cancels out. The eigenvalues
are then determined by the roots of the characteristic function

\[
X(\nu)=\nu-\sum_{k}\frac{|g|^{2}}{\nu-\omega_{k}}.\]

Since the $\omega_{k}$ are odd in $k$, this characteristic function
is odd as well:

\[
X(-\nu)=-X(\nu).\]

Thus there is always a root $\nu=0;$ the remaining roots appear as
pairs differing by a sign. It is useful to use this symmetry to combine
the $k$ and $-k$ terms in sum and write it as

\[
X(\nu)=\nu\left[1+\sum_{k=1}^{\Lambda}\frac{2|g|^{2}}{\omega_{k}^{2}-\nu^{2}}\right].\]

For each root $\nu_{\alpha}$ the eigenvector is given by

\[
U_{\alpha}^{k}=\frac{g}{\omega_{k}-\nu_{\alpha}}U_{\alpha}^{d}\]

This eigenvector will have length one if we set

\[
|U_{\alpha}^{d}|^{2}=\frac{1}{X'(\nu_{\alpha})}\]

since \[
X'(\nu)=1+\sum_{k}\frac{|g|^{2}}{\left(\nu-\omega_{k}\right)^{2}}.\]

Thus we have a $2\Lambda+1$ dimensional unitary matrix $u_{\alpha}^{K}$
that diagonalizes the effective hamiltonian:

\[
h=U\mathrm{diag}(\nu)U^{\dagger}.\]

\section{ The Ground State}

Since the static solution satisfies \[
[h(\Psi),\Psi]=0\]

it must also be diagonalized by $U$:

\[
\Psi=U\mathrm{diag}(\mu)U^{\dagger}\]

The eigenvalues $\mu_{\alpha}$of $\Psi$ are determined by the condition
that it describe the ground state of the system: the negative energy
states are occupied and the positive energy state is empty. The zero
energy state carries any electrons that are left over after these
assignments:

\[
\mu_{\alpha}=\left\{ \begin{array}{cc}
1,\quad & \nu_{\alpha}<0\\
\mu_{0},\quad & \nu_{\alpha}=0\\
0,\quad & \nu_{\alpha}>0\end{array}\right\}\]

The parameter $\mu_{0}$ is the total number of electrons divided
by $N$, modulo one. If $\mu_{0}=\frac{1}{2}$ we have just the right
number of electrons to have a ground state that is invariant under
the charge conjugation symmetry. It is useful to use instead a parameter
that measures the departure from this symmetric case:

\[
\xi=\mu_{0}-\frac{1}{2}.\]

If $N=2,$ an odd number of electrons correspond to $\xi=0$ and an
even number of electrons to $\xi=$$\frac{1}{2}$

Thus

\[
\mu_{\alpha}=\frac{1-\mathrm{sgn}(\nu_{\alpha})}{2}+\xi\delta_{\alpha,0}.\]
Of special interest are the elements

\[
\Psi_{d}^{k}=\sum_{\alpha}u_{\alpha}^{k}\mu_{\alpha}u_{d}^{*\alpha}\]

\[
=g\sum_{\alpha}\frac{\mu_{\alpha}}{\chi'(\nu_{\alpha})\left(\omega_{k}-\nu_{\alpha}\right)}\]

Separating out the zero-mode contribution that is not charge conjugation
invariant, the remaining sum can be written as a sum of residues:

\[
\Psi_{d}^{k}=\frac{g\xi}{\omega_{k}}\left[1+\sum_{m}\frac{|g|^{2}}{\omega_{m}^{2}}\right]^{-1}+g\frac{1}{2\pi i}\int_{D}\frac{dz}{X(z)\left(\omega_{k}-z\right)}\]
Here $D$ is a contour that starts at infinity a bit below the negative
real axis, goes through the origin (where a principal value is taken)
and then goes to infinity a bit above the negative real axis. The
only poles arise from the roots of $\chi(z)$ at which the residues
are given by the sum above, plus a principal value contribution from
the origin. We can now deform this contour to go along the imaginary
axis, and get the formula

\[
g\frac{1}{2\pi i}\int_{D}\frac{dz}{X(z)\left(\omega_{k}-z\right)}=\frac{g}{2\pi}\mathcal{P}\int_{-\infty}^{\infty}\frac{dy}{X(iy)(\omega_{k}-iy)}\]

\[
=\frac{g}{2\pi}\int_{0}^{\infty}\frac{dy}{X(iy)}\left[\frac{1}{\omega_{k}-iy}-\frac{1}{\omega_{k}+iy}\right]\]

\[
=\frac{g}{\pi}\mathcal{P}\int_{0}^{\infty}\frac{dy}{\omega_{k}^{2}+y^{2}}\frac{iy}{X(iy)}\]
Thus,

\[
\Psi_{d}^{k}=\frac{g\xi}{\omega_{k}}\ \left[1+\sum_{m}\frac{|g|^{2}}{\omega_{m}^{2}}\right]^{-1}\ +\frac{g}{\pi}\mathcal{P}\int_{0}^{\infty}\frac{dy}{\left[\omega_{k}^{2}+y^{2}\right]\left[1+2|g|^{2}\Sigma(y)\right]}\]

where

\[
\Sigma(y)=\sum_{m>0}\frac{1}{y^{2}+\omega_{m}^{2}}.\]

This sum converges even as $\Lambda\to\infty$; also the $\Sigma(y)\sim\frac{\pi}{2cy}$
for large $y$.

Setting

\[
y=|\omega_{k}|x,\]

we get for the part even in $k$,

\[
\Psi_{+d}^{k}=\frac{g}{\pi|\omega_{k}|}\int_{0}^{\infty}\frac{dx}{\left[1+x^{2}\right]\left[1+2|g|^{2}\Sigma(|\omega_{k}|x)\right]}\ \]

It follows that

\[
\lim_{|k|\to\infty}|\omega_{k}|\Psi_{+d}^{k}=\frac{g}{2}.\]

This will be useful for renormalization.

It is possible to get evaluate the sums when the spectrum is exactly
(not just asymptotically) linear $\omega_{k}=ck$ .

\[
\Sigma(y)=\frac{-c+\pi y\text{coth}\left[\frac{\pi y}{c}\right]}{2cy^{2}}\]

With

\[
g'=\frac{g}{c}\]

we have

\[
\Psi_{d}^{k}=\frac{g'}{\left[1+\zeta(2)|g'|^{2}\right]}\frac{\xi}{k}+\frac{g'}{\pi}\int_{0}^{\infty}\frac{a^{2}da}{\left[k^{2}+a^{2}\right]\left[a^{2}+|g'|^{2}\left(\pi a\mathrm{coth}(\pi a)-1\right)\right]}\]

\begin{figure}

\caption{\protect\includegraphics{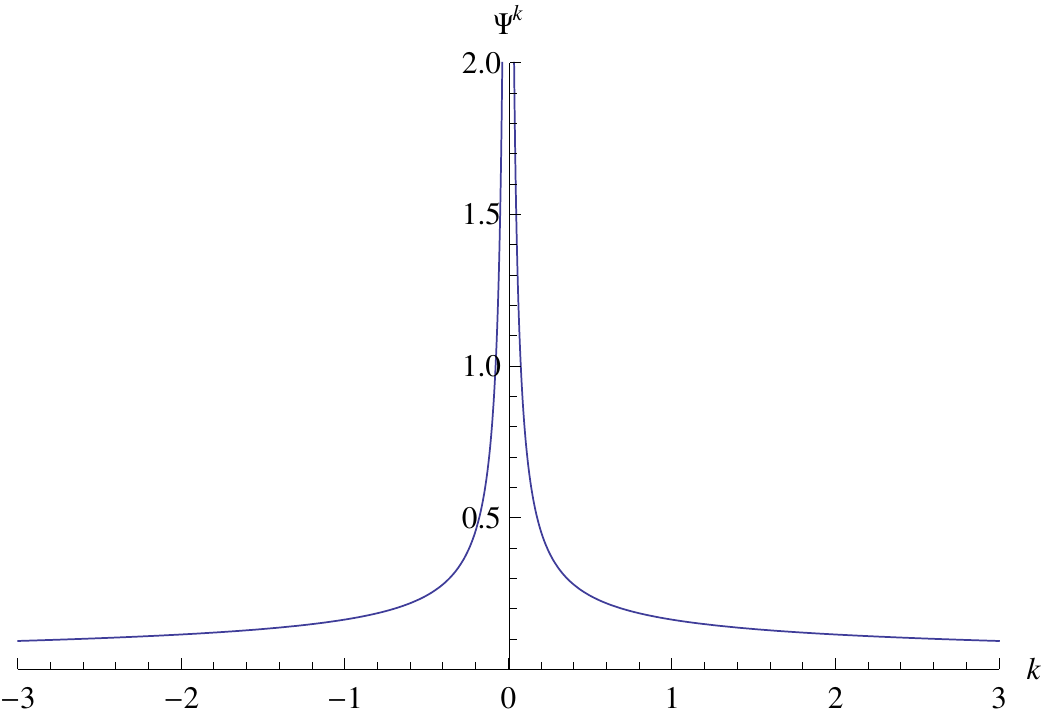}}

The impurity-electron condensate when $\xi=0.$ Note the symmetry
$k\to-k.$

\end{figure}

\begin{figure}

\caption{\protect\includegraphics{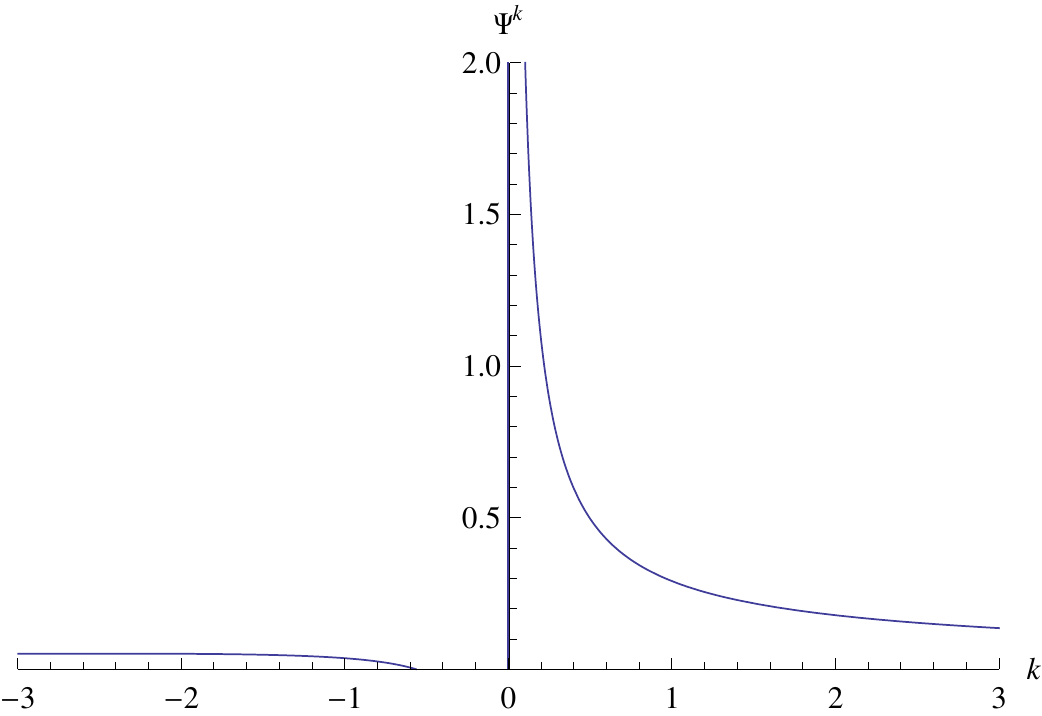}}

The impurity-electron condensate when $\xi=0.5.$ The symmetry $k\to-k$
is broken.
\end{figure}

\section{Renormalization}

So far we have studied the problem for a fixed value of $\Lambda.$
If we look back at the expressions in the last sections, we see that
the limit $\Lambda\to\infty$ is convergent for $\Sigma,X,\Psi$ and
$h$, as long as $g$ is kept fixed. Since

\[
g=J\sum_{k}\Psi_{+d}^{k}\]
and the sum is log divergent, it follows that $J\sim\frac{1}{\log\Lambda}.$
More precisely, (remembering that only the even part of $\Psi_{d}^{k}$
contributes to the sum over $k$)

\[
\lim_{\Lambda\to\infty}J(\Lambda)\sum_{k=1}^{\Lambda}\frac{1}{\omega_{k}}=1.\]
This is asymptotic freedom. During renormalization we trade the divergent
constant $J^{-1}$ for $g$ , which remains finite as $\Lambda\to\infty.$
In detail,

\[
J^{-1}(\Lambda,g)=\sum_{k=1}^{\Lambda}\frac{2}{\pi}\int_{0}^{\infty}\frac{dy}{\left[\omega_{k}^{2}+y^{2}\right]\left[1+2|g|^{2}\Sigma(y)\right]}.\]
The dependence on $g$ is sub-leading order in $\Lambda$.

It is worth noting that $g$ is a complex-valued parameter, although
the original anti-ferromagnetic coupling $J$ is real. The symmetry
generated by the conserved quantity $\Phi_{d}^{d}$ (the number of
electrons occupying the impurity site) is spontaneously broken, as
it corresponds to the phase of $g.$ If there is a lattice of impurities,
this would become a translation invariant field that breaks the gauge
invariance of electromagnetism spontaneously: a possible mechanism for superconductivity
in heavy fermion systems\cite{HeavyFermion}.

\section{The Renormalized Theory}

The sum defining the characteristic function is convergent in the
limit $\Lambda\to\infty:$

\[
\chi(\nu)=\nu\left[1+\sum_{k=1}^{\infty}\frac{2|g|^{2}}{\omega_{k}^{2}-\nu^{2}}\right].\]

Removing the overall factor of $\nu$(which just gives the obvious
root at $\nu=0$), we get the function

\[
\chi_{1}(\nu)=1+\sum_{k=1}^{\infty}\frac{2|g|^{2}}{\omega_{k}^{2}-\nu^{2}}.\]

In each interval $[\omega_{k},\omega_{k+1}]$ (with $\omega_{k}>0)$
$\chi_{1}(\nu)$ increases from $-\infty$ to $\infty$ monotonically.
Thus it has exactly one root $\nu_{\alpha}$ in each such interval.

\begin{figure}

\caption{\protect\includegraphics{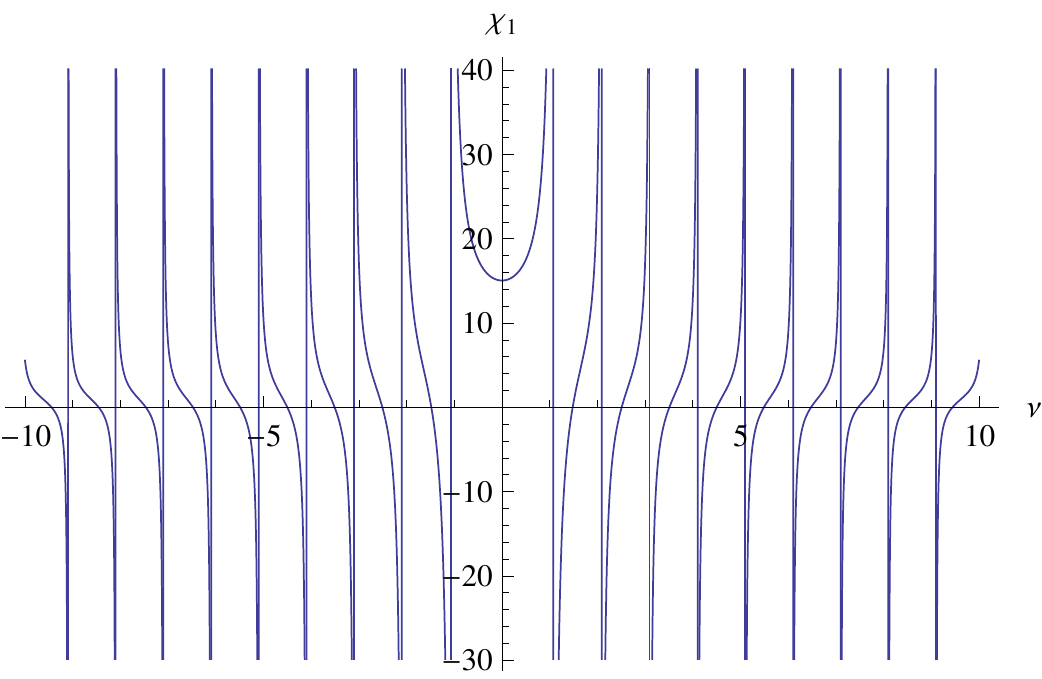}}

The characteristic function $\chi_{1}(\nu)$ for a nearly dispersion
relation.
\end{figure}

The ground state $\Psi$ and the effective hamiltonian $h=h(\Psi)$
of (\ref{eq:h}) also make sense as $\Lambda\to\infty$keeping $g$
fixed.

We can use the departure from the ground state as the dynamical variable
:

\[
\Psi=\Psi+\phi.\]

In addition, it is natural to rotate to the basis in which $h(\Psi)$
is diagonal.That is, put

\[
\phi_{L}^{K}=\phi_{\beta}^{\alpha}U_{\alpha}^{K}U_{L}^{*\beta}\]

and use the components $\phi_{\beta}^{\alpha}$as our dynamical variables.
Then the effective hamiltonian becomes

\[
h_{\beta}^{\alpha}(\phi)=\nu_{\alpha}\delta_{\beta}^{\alpha}-\left[r(\phi)U_{\alpha}^{d}U_{\bullet}^{*\beta}+h.c\right]\]

where

\[
r(\phi)=J\sum_{\gamma,\delta}\phi_{\delta}^{\gamma}U_{d}^{*\delta}U_{\gamma}^{\bullet}.\]

Now,

\[
U_{\alpha}^{\bullet}=-\nu_{\alpha}\frac{U_{\alpha}^{d}}{g^{*}}\]

\[
r(\phi)=-J\sum_{\alpha,\beta}\nu_{\alpha}\phi_{\beta}^{\alpha}\frac{U_{d}^{*\beta}U_{\alpha}^{d}}{g^{*}}=-\frac{J}{g^{*}}\sum_{\alpha,\beta}\frac{\nu_{\alpha}\phi_{\beta}^{\alpha}}{\sqrt{X'(\nu_{\alpha})X'(\nu_{\beta})}}\]

Because of the overall factor of $J,$ this will vanish unless the
sum in $\alpha,\beta$ diverges. Thus, for those $\phi_{\beta}^{\alpha}$with
just a \emph{finite number of non-zero entries}

\[
h_{\beta}^{\alpha}(\phi)=\nu_{\alpha}\delta_{\beta}^{\alpha}\]

Such finite rank configurations satisfy the\textbf{\emph{ }}\emph{linear}
evolution equation equation

\[
-i\frac{d\phi_{\beta}^{\alpha}}{dt}=[\nu_{\alpha}-\nu_{\beta}]\phi_{\beta}^{\alpha}.\]

This subset is closed under time evolution. The corresponding quantum
states, are free quasi-particles.

\section{New Dynamical Variables}

The dynamical variables $\phi_{\beta}^{\alpha}$ obtained after subtracting
the static solution $\Psi$, and passing to the basis diagonalizing
$h(\Psi)$, satisfy the Poisson brackets

\begin{equation}
-i\left\{ \phi_{\beta}^{\alpha},\phi_{\delta}^{\gamma}\right\} =\delta_{\beta}^{\gamma}\phi_{\delta}^{\alpha}-\delta_{\delta}^{\alpha}\phi_{\beta}^{\gamma}+\left(\mu_{\alpha}-\mu_{\gamma}\right)\delta_{\delta}^{\alpha}\delta_{\beta}^{\gamma}\label{eq:NewPB}\end{equation}

(Recall that $\mu_{\alpha}$are the eigenvalues of the static solution
$\Psi$.) This is the central extension of the unitary Lie algebra,
defined for example, in the book by Pressley-Segal\cite{PressleySegal}.
If only a finite number of the $\phi_{\beta}^{\alpha}$ are non-zero,
we can supplement this with an element describing time evolution

\[
-i\left\{ h,\phi_{\beta}^{\alpha}\right\} =[e_{\alpha}-e_{\beta}]\phi_{\beta}^{\alpha}.\]

All the effects of the impurity-electron interaction are contained
in the shift of the energies from $\omega_{k}$ to $\nu_{\alpha}$
and in the occupation numbers $\mu_{\alpha}.$ The unitary transformation
$U_{\alpha}^{K}$ relates the new degrees of freedom to the old. In
the continuum limit, this can be expressed as a scattering phase shift
of the electrons.

\section{The Case of Finite N}

Now we are ready to return to the case of finite $N.$ Since $\frac{1}{N}$
plays a role analogous to $\hbar,$ this amounts to quantizing the
Poisson brackets (\ref{eq:NewPB}). That is, find operators that satisfy
these commutation relations

\begin{equation}
\left[\hat{\phi}_{\beta}^{\alpha},\hat{\phi}_{\delta}^{\gamma}\right]=\frac{1}{N}\left(\delta_{\beta}^{\gamma}\hat{\phi}_{\delta}^{\alpha}-\delta_{\delta}^{\alpha}\hat{\phi}_{\beta}^{\gamma}+\left[\mu_{\alpha}-\mu_{\gamma}\right]\delta_{\delta}^{\alpha}\delta_{\beta}^{\gamma}\right)\label{eq:NewCR}\end{equation}

The representation of interest is

\[
\hat{\phi}_{\beta}^{\alpha}=\frac{1}{N}:a^{\dagger\alpha\sigma}a_{\beta\sigma}:.\]

where $a,a^{\dagger}$ are fermionic operators and the  normal
ordering is with respect to the Dirac vacuum of the energies $e_{\alpha}$:

\[
a^{\dagger\alpha}\mid0\rangle=0,\quad e_{\alpha}<0\]

\[
a_{\alpha}\mid0\rangle=0,\quad e_{\alpha}>0.\]

The hamiltonian just describes quasi-particles with these energies:

\[
\hat{H}=\frac{1}{N}\sum_{\alpha}\nu_{\alpha}:a^{\dagger\alpha\sigma}a_{\alpha\sigma}:\]

We get free particles only because we ignored terms in the hamiltonian
that are not divergent. If we add UV finite interactions to the hamiltonian
in addition, we get a Fermi liquid.

\section{Acknowledgement}

I thank R. Henderson (in 1994), P. Jacquot, A. Jordan and J. Polchinski
for discussions. This work was supported in part by a grant from the
US Department of Energy under contract DE-FG02-91ER40685.


\begin{thebibliography}{16}
\bibitem{Huang}K. Huang, {}``Quarks, Leptons and Gauge Fields\textquotedblright{},
World Scientific, Singapore (1982).

\bibitem{AdSCFT}O.Aharony, S. S. Gubser, J. Maldacena, H. Ooguri
and Y.Oz Physics Reports 323: 183-386(2000).

\bibitem{QYMT} A.Jaffe and E. Witten,\emph{Quantum Yang-Mills Theory
},http://www.claymath.org/millennium/Yang-Mills\_Theory/Official\_Problem\_Description.pdf

\bibitem{Hewson} A. C. Hewson, {}``The Kondo Problem to Heavy Fermions'',
Cambridge University Press (1997)

\bibitem{Wilson}K. G. Wilson, Rev. Mod. Phys. 47, 773 - 840 (1975)

\bibitem{Andrei} N. Andrei Phys. Rev. Lett., 45, 379 (1980).

\bibitem{Wiegmann}P. B. Wiegmann, Soviet Physics JETP Letters, 31,
392(1980).

\bibitem{Nozieres} P. Nozieres Jour. Low Temp. Phys., 17, 31 (1974).

\bibitem{AffleckLudwig}I. Affleck and A. W. Ludwig Nucl. Phys. B360
641(1991).

\bibitem{Neoclassical} S. G. Rajeev, {}``New Classical Limits Of
Quantum Theories'',in {}``Infinite Dimensional Groups and Manifolds''
Edited by T. Wurzbacher Berlin, New York (Walter de Gruyter) (2004)
Pages 213\textendash{}248;{[}arxiv:hep-th/0210179{]}

\bibitem{Mora}C. Mora, Phys. Rev. B80, 125304 (2009) and references
therein

\bibitem{QCD2} S. G. Rajeev, Int.J.Mod.Phys. A9 (1994) 5583-5624;
{[}arxiv:hep-th/9401115{]}

\bibitem{HendersonRajeev}R. J. Henderson and S. G. Rajeev,Int.J.Mod.Phys.
A10 (1995) 3765-3780{[}arXiv:hep-th/9501080{]}

\bibitem{QDot}S. M. Cronenwett, T. H. Oosterkamp and L. P. Kouwenhoven,
Science, 281, 540 (1998)

\bibitem{PressleySegal}A. Pressley and G. segal, {}``Loop Groups''
, Oxford Universoty Press (1988)

\bibitem{HeavyFermion}H. B. Radousky, {}``Magnetism in Heavy Fermion
Systems'', World Scientific (2000).
\end{thebibliography}
\end{document}